\documentclass{aa}
\usepackage[varg]{txfonts}
\pdfoutput=1
\bibpunct{(}{)}{;}{a}{}{,} 

\usepackage[colorlinks,linkcolor=blue,
           anchorcolor=blue,citecolor=blue
           ]{hyperref}
\usepackage{natbib}
\usepackage{amsmath}
\usepackage{amssymb}
\usepackage{url}
\usepackage{longtable}
\usepackage{multirow}

\usepackage{graphicx}   
\usepackage{epsfig}
\usepackage{bm}

\newcommand{\EqAA}{{Eq.}}

\newcommand{\Fig}{{Figure}}
\newcommand{\Figs}{{Figures}}
\newcommand{\FigAA}{{Fig.}}
\newcommand{\FigsAA}{{Figs.}}
\newcommand{\Sec}{{Sect.}}

\newcommand{\RNum}[1]{\uppercase\expandafter{\romannumeral #1\relax}}

\begin{document}
  \title{Numerical Simulation of a Fundamental Mechanism of Solar Eruption with Different Magnetic Flux Distributions}


 \author{Xinkai Bian\inst{\ref{inst1}}
    \and Chaowei Jiang\inst{\ref{inst1}}
	\and Xueshang Feng\inst{\ref{inst1}}
	\and Pingbing Zuo\inst{\ref{inst1}}
	\and Yi Wang\inst{\ref{inst1}}
	\and Xinyi Wang\inst{\ref{inst2}} }	
  \institute{Institute of Space Science and Applied Technology, Harbin Institute of Technology, Shenzhen 518055, China\\
  	 \email{chaowei@hit.edu.cn} \label{inst1}
  \and SIGMA Weather Group, State Key Laboratory for Space Weather, National Space Science Center, Chinese Academy of Sciences,
  Beijing 100190, China \label{inst2} }

	
\abstract{Solar eruptions are explosive release of coronal magnetic field energy as manifested in solar flares and coronal mass ejection. Observations have shown that the core of eruption-productive regions are often a sheared magnetic arcade, i.e., a single bipolar configuration, and, particularly, the corresponding magnetic polarities at the photosphere are elongated along a strong-gradient polarity inversion line (PIL). It remains unclear what mechanism triggers the eruption in a single bipolar field and why the one with a strong PIL is eruption-productive. Recently, using high accuracy simulations, we have established a fundamental mechanism of solar eruption initiation that a bipolar field as driven by quasi-static shearing motion at the photosphere can form an internal current sheet, and then fast magnetic reconnection triggers and drives the eruption. Here we investigate the behavior of the fundamental mechanism with different photospheric magnetic flux distributions, i.e., magnetograms, by combining theoretical analysis and numerical simulation. Our study shows that the bipolar fields of different magnetograms, as sheared continually, all exhibit similar evolutions from the slow storage to fast release of magnetic energy in accordance with the fundamental mechanism, which demonstrates the robustness of the mechanism. We further found that the magnetograms with stronger PIL produce larger eruptions, and the key reason is that the sheared bipolar fields with stronger PIL can achieve more non-potentiality, and their internal current sheet can form at a lower height and with a larger current density, by which the reconnection can be more efficient. This also provides a viable trigger mechanism for the observed eruptions in active region with strong PIL.}

\keywords{Sun: magnetic fields -- 
		Methods: numerical -- 
		Sun: corona --
		Magnetohydrodynamic (MHD)}
\titlerunning{Numerical Simulation of Solar Eruption}
\authorrunning{Bian et al.}
\maketitle

\section{Introduction}
\label{sec:intro}

Solar flares and coronal mass ejections (CMEs) are violent eruptions on the Sun, and essentially they are manifestation of the explosive release of magnetic energy in the solar corona. It is commonly believed that the solar magnetic fields are generated at the base of the convection zone, often in the form of thin, intense flux tubes, and emerge outwards slowly, eventually into the corona through the solar surface (i.e., the photosphere). Thereafter, the coronal magnetic fields will be continuously, but rather slowly, dragged at their footpoints by the photospheric surface motions, often in a organized form such as shearing and rotational flows, which is also inherent to the flux emergence process. During this process, magnetic free energy in the corona gradually accumulates, and the magnetic field configuration is often built up to a highly stressed one of S shape (e.g., the coronal sigmoids frequently observed in X-ray and EUV images) prior to an eruption.
Before the eruption onset, the coronal system is in an equilibrium state, in which the outward magnetic pressure of the low-lying, strongly stressed flux is balanced by the inward magnetic tension of the overlying, mostly un-sheared flux. At a critical point, the eruption suddenly begins with a catastrophic disruption of this force balance, during which the free magnetic energy is rapidly converted into impulsive heating and fast acceleration of the plasma.

It remains an open question how solar eruptions are initiated, namely, how the force balance before eruption is suddenly destroyed and what drives the eruption, for which many theories have been proposed~\citep{Forbes2006, Shibata2011, ChenP2011, Schmieder2013, Aulanier2014, Janvier2015}. The existing theories are often divided into two categories, one is based on the ideal magnetohydrodynamic (MHD) instability, and the other on magnetic reconnection. The first category generally requires the pre-existence of magnetic flux rope (MFR), a group of twisted magnetic field lines that wind tightly about a common axis, and the ideal instabilities of the MFR, such as kink instability and torus instability can initiate eruptions~\citep{Kliem2006, Torok2005, Fan2007, Aulanier2010, Amari2018}. In the second category, the most frequently mentioned models are the breakout model and the tether-cutting reconnection model. The breakout model requires a quadrupolar magnetic configuration in which a magnetic null point situates above the core of sheared magnetic flux. It is proposed that the reconnection at the null point removes the overlying restraining flux to trigger an eruption~\citep{Antiochos1999,Aulanier2000,Lynch2008,Wyper2017}. The tether-cutting reconnection model relies on only a single sheared arcade, i.e., a bipolar magnetic field. With increasing of the magnetic shear, a current sheet (CS) will be formed slowly at a low altitude above the photospheric magnetic PIL. Initially, the magnetic reconnection at that CS slowly reduces the downward magnetic tension force by ``cutting the magnetic tethers'', and then the upward magnetic pressure force is unleashed and pushes up the flux to rise, which in turn enhances the tether-cutting reconnection. After a short interval, the process becomes runaway, which triggers the eruption, and the fast rising flux stretches up the surrounding envelope magnetic field, forming a new elongated CS above the PIL.
The magnetic reconnection of the newly-formed CS further speeds up the eruption to form a CME ~\citep{Moore1980,Moore1992,Moore2001}. Compared with other models, the tether-cutting scenario is the simplest one in terms of magnetic topology, since it relies on a single magnetic arcade (corresponding to a pair of opposite polarities at the photosphere) without any additional special topology, such as null point and MFR. However, unlike other models that have been extensively realized in numerical 3D MHD simulations~\footnote{Note that the aforementioned references cited for the ideal MHD instability and the breakout models are mostly based on numerical simulations.}, the tether-cutting model has not yet been validated in any 3D MHD simulations, and thus remains a conjectural ``cartoon.''

Actually, early simulations in 2D or translational invariant geometries ~\citep{mikic_disruption_1994,Amari1996,Choe1996} show that by continuous shearing of its footpoints, a single magnetic arcade asymptotically approaches an open state containing a CS, which is consistent with the Aly-Sturrock conjecture ~\citep{aly_how_1991,sturrock_maximum_1991}. When one takes into account finite resistivity, the system experiences a global disruption once reconnection sets in at the CS, which, in particular, begins at the point with the largest current density in the CS. Such an simple and efficient mechanism of eruption initiation is only recently established in fully 3D by~\citet{jiang_fundamental_2021} with an ultra-high accuracy MHD simulation. That simulation is initialized with a bipolar potential field. Through surface shearing motion along the PIL, a vertical CS forms quasi-statically above the PIL. Once the CS is sufficiently thin such that ideal MHD is broken down, reconnection sets in and instantly triggers the eruption. The simulation shows that the reconnection not only cuts the magnetic tethers, but also results in strong upward tension force, and it is the latter that plays the key role in driving the eruption, that is, the slingshot effect of the reconnection drives mainly the eruption. We note that this mechanism is different from the tether-cutting model in twofold. Firstly, the tether-cutting model proposed that the reconnection only plays a role of cutting the confinement of the field lines, and it is the unleashed magnetic pressure that drives the eruption. Secondly, the tether-cutting model assumed that before the onset of the eruption (i.e., the start of impulsive phase of eruption), there is a relatively long phase of slow reconnection with tens of minutes to a few hours, which gradually cuts the tethers, until a ``global instability'' occurs can then the eruption begin~\citep{Moore2001}. Such slow reconnection is not runaway, and it does not exist in~\citet{jiang_fundamental_2021}'s simulation. In this sense, the model as demonstrated by \citet{jiang_fundamental_2021} stands alone with the original tether-cutting model, and hereafter, we called this mechanism as the BASIC model, where the acronym "BASIC" refers to the key ingredients as involved in the mechanism, that is, a Bipolar magnetic Arcade as sheared evolves quasi-Statically and forms Internally a Current sheet.

As is demonstrated that the solar eruption can be initiated from a single bipolar field by the BASIC mechanism, a natural question arises; how does the mechanism operate differently with different flux distributions of bipolar field on the photosphere? Since \citet{jiang_fundamental_2021} carried out numerical experiments for only one set of magnetic flux distribution on the bottom surface, in this paper we will investigate the BASIC mechanism with different magnetic flux distributions. In particular, we aim to know what kind of flux distribution of bipolar field on the photosphere is favorable for producing major eruptions and why. This study is motivated by a well known fact from observations that major solar eruptions occur predominantly in source regions having a pronounced PIL with both an elongated distribution of flux along it and a strong gradient of field across it~\citep{schrijver_characteristic_2007, Toriumi2019}. Such a PIL is mostly often found in a particular type of sunspot group, the $\delta$ sunspots, which are highly flare-productive as first found by \citet{kunzel_flare-haufigkeit_1959}. By examining the magnetic properties of regions associated with almost $300$ M- and X-class flare, \citet{schrijver_characteristic_2007} found that the magnetic flux distribution in the flare site of these regions is often of bipolar configuration with a characteristic pattern: a relatively elongated, strong-gradient PIL with strong magnetic shear. Such a pattern is even employed in developing empirical models of flare forecast by calculating the length of the high-gradient PIL~\citep{Falconer2002,falconer_measure_2003}. So, why a strong-gradient and long PIL, or simply a strong PIL, is more favorable for producing eruption? To the best of our knowledge, the physical reason behind the correlation of this property of PIL and the eruption-productiveness has never been explained explicitly. This paper is devoted to answer this question based on the BASIC mechanism, by performing a series of 3D MHD simulations similar to~\citet{jiang_fundamental_2021}'s but with different photospheric flux distributions (or, magnetograms) for a comparative study. Some of magnetograms have strong PIL while others have weak PIL (i.e., short and weak-gradient one). As will be seen, nearly all the simulations follow the BASIC scenario of quasi-static formation of CS and triggering of eruption by reconnection, which demonstrates the robustness of the BASIC mechanism, and strikingly, the magnitude of the eruption is highly dependent on the strength of the PIL.

The paper is organized as follows. In \Sec~\ref{sec:magnetogram}, we define the magnetograms for the bipolar fields with different magnetic flux distributions. Before showing the simulation results, in \Sec~\ref{sec:magnetic field analyze} we analyze the key parameters associated with the non-potentiality and CS of the open field configuration corresponding to these different bipolar fields, since the BASIC mechanism is closely linked to the open field. Then we compare the results of MHD simulations for the different bipolar fields in Sects. \ref{sec:model} and \ref{sec:results}, and give our conclusion and discussions in \Sec~\ref{sec:conclusions}.

\section{Numerical Experiments}
\label{sec:Numerical Experiments}

\subsection{Setting of magnetograms}
\label{sec:magnetogram}
\begin{figure*}[!htbp]
	\centering
	\includegraphics[width=0.6\textwidth]{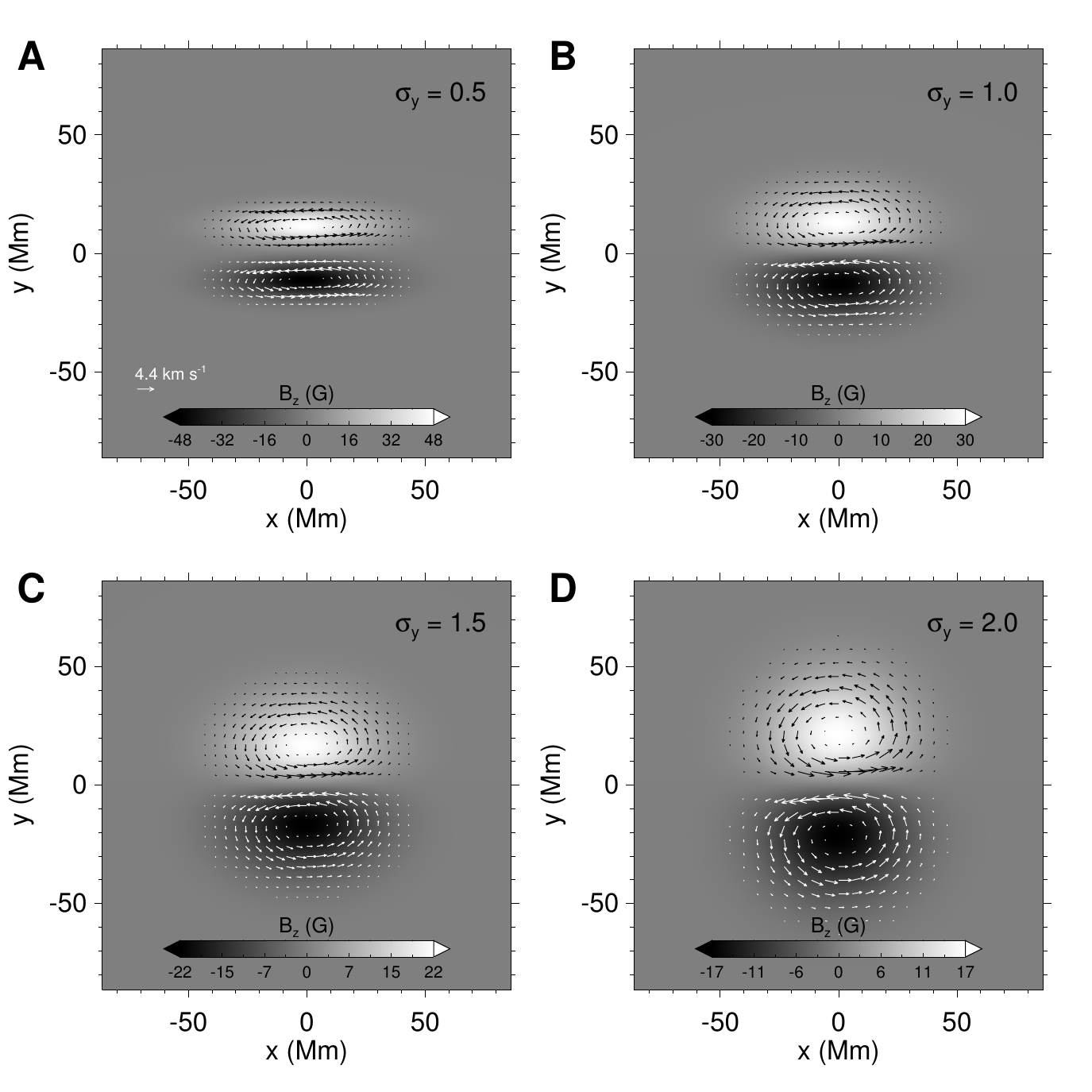}
	\caption{Magnetic flux distribution and surface rotation flow at the bottom surface (i.e., $z=0$). The background is color-coded by the vertical magnetic component $B_{z}$, and the vectors show the rotation flow. The four panels~\textbf{A}, \textbf{B}, \textbf{C} and \textbf{D} refer to magnetogram with $\sigma_y = 0.5$, $1.0$, $1.5$ and $2.0$ respectively, while $ \sigma_x = 2.0 $ and $ y_c = 0.8 $ are fixed. Note that the values of $\sigma_x$, $\sigma_y$ and $y_c$ are normalized by a unit length of $14.4$ Mm.}
	\label{fig:sigy_bz_four_figure}
\end{figure*}

\begin{figure*}[!htbp]
	\centering
	\includegraphics[width=0.6\textwidth]{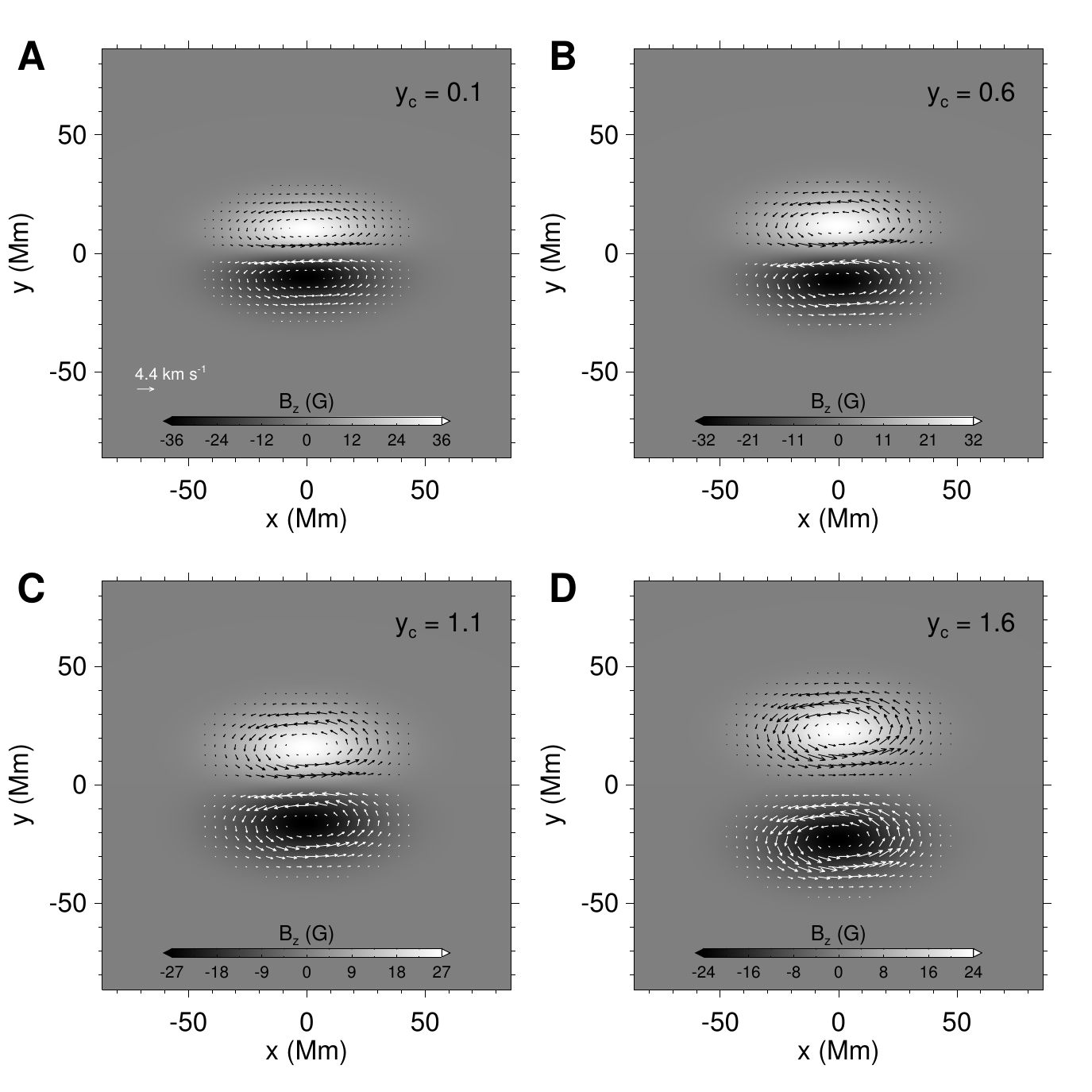}
	\caption{Same as \FigAA~\ref{fig:sigy_bz_four_figure} but with different values of $y_c = 0.1$, $0.6$, $1.1$ and $1.6$ respectively, while $\sigma_x = 2.0$ and $\sigma_y = 1.0$ are fixed.}
	\label{fig:yc_bz_four_figure}
\end{figure*}

Following \citet{Amari2003A} and \citet{jiang_fundamental_2021}, we model the photospheric magnetogram of a bipolar field by the composition of two Gaussian functions,
\begin{equation}\label{eq:magnetogram}
	\begin{split}
		B_z(x,y,0) = B_0 e^{-x^2/\sigma_x^2}(e^{-(y^2-y_c^2)/\sigma_y^2}-e^{-(y^2+y_c^2)/\sigma_y^2}),
	\end{split}
\end{equation}
where $\sigma_x$ and $\sigma_y$ control the extents of the magnetic flux distribution in \textit{x} and \textit{y} directions, respectively, and $y_c$ control the distance between the two magnetic polarities in the \textit{y} direction. By adjusting these controlling parameters, we obtain different magnetograms with different flux distributions, some of which have long and strong-gradient PIL while other have short and weak ones. For example, by increasing $\sigma_y$ but fixing $\sigma_x$ and $y_c$, the shape of the polarities gradually changes from flat to round and the strong-gradient PIL becomes shorter, as can be seen in \FigAA ~\ref{fig:sigy_bz_four_figure} with four values of $\sigma_y$. In another way, by increasing $y_c$ but fixing $\sigma_x$ and $\sigma_y$, the two opposite polarities become further away from each other and thus the field gradient across the PIL decreases, as shown in \FigAA~\ref{fig:yc_bz_four_figure} for four values of $y_c$. For a reasonable comparison, we use the same value for the total unsigned magnetic flux of all the different magnetograms, as such the different magnetograms only differ in the distribution of the same amount of magnetic flux. Therefore, the larger the area of magnetic polarity is, the weaker the average magnetic field strength is, and vice versa. For the specific values of the parameters, we choose $[-10,10] \times [-10,10]$ with length unit of $14.4$~Mm as our computational domain. The values for the parameters as shown in \FigsAA~\ref{fig:sigy_bz_four_figure} and \ref{fig:yc_bz_four_figure} are given in their captions.

\subsection{The open field}
\label{sec:magnetic field analyze}

\begin{figure*}[!htbp]
	\centering
	\includegraphics[width=0.6\textwidth]{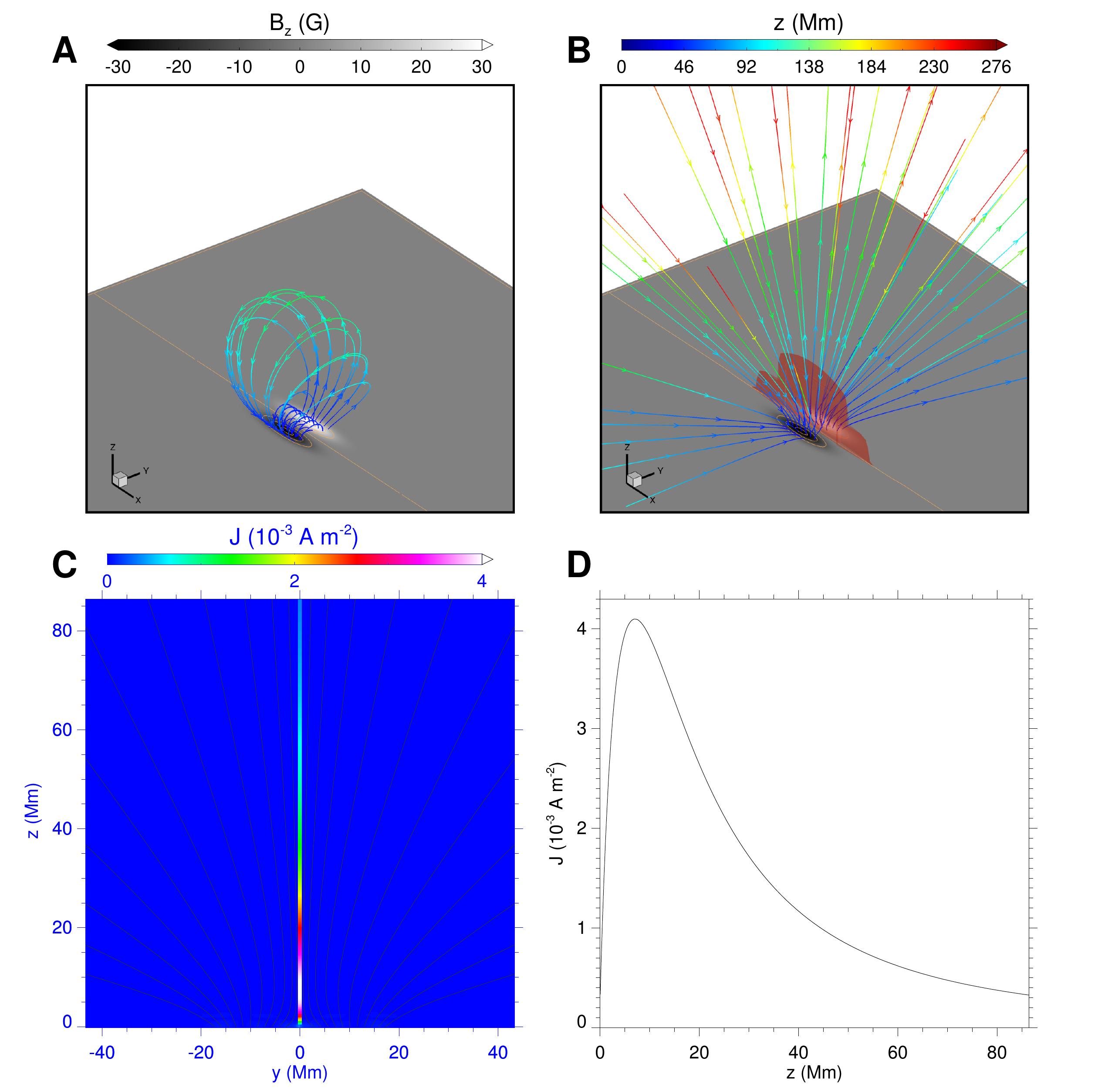}
	\caption{The potential field and the open field with the same magnetic flux distribution on the bottom surface, and the distribution of CS in the open field. \textbf{A} 3D prospective view of the potential field lines. The colored thick lines represent magnetic field lines and the colors denote the height. The background shows the magnetic flux distribution on the bottom boundary. \textbf{B} 3D prospective view of the fully opened field lines, with same footpoints shown in panel \textbf{A}. The red iso-surface represents the CS which current density $J=0.34 \times 10^{-3}$ A m$^{-2}$. \textbf{C} Current distribution and magnetic field lines on the central cross section (i.e., the $x=0$ plane). \textbf{D} Profile of current density along $z$ axis. Note that here with finite grid resolution, the CS has a finite thickness of 360 km. Thus the current density is not infinite.}
	\label{fig:open_slice}
\end{figure*}

The BASIC mechanism is closely linked to the fully open field, because by ideally shearing a magnetic arcade, it asymptotically approaches the open field state. In \FigAA~\ref{fig:open_slice}, we show an example of the open field (and compared with the corresponding potential field) for a bipolar magnetogram. The procedure of the open field computation is as follows: first, we calculate the potential field $\textbf{B}_{\rm pot,u}(x,y,z)$ based on a unipolar magnetogram defined by $B_{z,\rm u}(x,y,0)={\rm sign}(y)B_{z}(x,y,0)$. The potential field is solved using the Green's function method. Thus, all the field lines are open, going outwards from the bottom boundary to infinity. Then, we reverse the direction of the field lines that root in the original negative polarity, and in our case this can be simply done by defining $\textbf{B}_{\rm open}(x,y,z)={\rm sign}(y)\textbf{B}_{\rm pot,u}(x,y,z)$ according to the symmetry of the field. In this way, all the field is still open and current free, except at the interface (i.e., the $y=0$ plane) between the inverse-directed field lines, which forms the CS. Therefore this is the open field corresponding to the bipolar flux distribution. 

The open field has important implications on the content of energy that a sheared bipolar field of the same magnetogram can store, as well as the intensity of CS that the sheared bipolar field can form. According to the Aly-Sturrock conjecture, the energy of such open field is the upper limit of the energy of all possible force-free fields with a given magnetic flux distribution on the bottom and a simply connected topology~\citep{aly_how_1991,sturrock_maximum_1991}. Therefore, the upper limit of free magnetic energy that can be reached in a sheared arcade is the open field energy subtracted by the corresponding potential energy of the same magnetogram,
\begin{equation}\label{eq:free_limit}
	\begin{split}
		E_{\rm uf} = E_{\rm open} - E_{\rm pot},
	\end{split}
\end{equation}
where $E_{\rm uf}$ denotes the upper limit of the free energy. If using the ratio of free magnetic energy to the potential energy as a measure of the non-potentiality of the field, namely $N = E_{\rm free}/E_{\rm pot}$, the upper limit of the non-potentiality of a sheared arcade can achieve is $N_{\rm max} = E_{\rm uf}/E_{\rm pot}$. The degree of non-potentiality $N$ of a field has been suggested as a critical factor in producing major eruptions~\citep{moore_limit_2012, SunX2015}, and thus $N_{\rm max}$ is an important indicator for the eruption capability of a magnetogram. For example, \citet{moore_limit_2012} studied a large number of active regions and found that ``there is a sharp upper limit to the free energy the field can hold that increases with the amount of magnetic field in the active region, the active region's magnetic flux content, and that most active regions are near this limit when their field explodes in a coronal mass ejection/flare eruption.'' In particular, using a proxy of magnetic shear for the free energy, they concluded that the non-potentiality $N$ is on the order of one for the active region's core field, i.e., field rooted around the flare PIL, when the field is close to eruption. Therefore, for a sheared bipolar field that can produce major eruption, it should have its upper limit of non-potentiality somewhat close or above one, i.e., $N_{\rm max} \gtrsim 1$, and by calculating this parameter for a given magnetogram, one can charge whether it is capable of generating large eruption or not.

The potential field $\textbf{B}_{\rm pot}$ and the open field $\textbf{B}_{\rm open}$ are uniquely defined by
\begin{equation}\label{eq:boundary_bz}
	\begin{split}
		B_{z|\rm pot}(x,y,0) = B_{z|\rm open}(x,y,0) = B_{z}(x,y,0)
	\end{split}
\end{equation}
and an asymptotic decay at infinity. These fields have energies given, respectively, by the standard relations~\citep[e.g.,][]{Amari2003A}
\begin{equation}\label{eq:energy_pot_open}
	\begin{split}
		E_{\rm pot} &= \dfrac{1}{16\pi^{2}} \int_{S\times S'} \dfrac{B_{z}(x,y,0)B_{z}(x',y',0)}{|\textit{\textbf{r}}-\textit{\textbf{r}}'|}{\rm d}s{\rm d}s', \\
		E_{\rm open} &= \dfrac{1}{16\pi^{2}} \int_{S\times S'} \dfrac{|B_{z}(x,y,0)B_{z}(x',y',0)|}{|\textit{\textbf{r}}-\textit{\textbf{r}}'|}{\rm d}s{\rm d}s'.
	\end{split}
\end{equation}
The upper limit of free magnetic energy can be obtained by \EqAA~(\ref{eq:free_limit}), and then the non-potentiality $N_{\rm max}$ of each magnetogram can be calculated. By the way, the magnetic field (potential field and open field) in volume can be obtained from the bottom magnetogram, and then the magnetic energy can be obtained by integration $E=\frac{1}{8\pi}\int B^{2} {\rm d}V$. However computing the magnetic field in the full volume is very time consuming. For example, with the Green's function method, the computing time scales with the grid number as $N^5$ (assuming the volume is a cube and the length of each side is $N$). That is why we choose to use \EqAA~(\ref{eq:energy_pot_open}) which is much faster in calculation, since the magnetic energy can be obtained directly from the magnetogram without knowing the magnetic field in the full volume. The magnetic energy obtained by these two methods is basically the same. For the magnetogram in \FigAA~\ref{fig:sigy_bz_four_figure}A, the potential field energy and open field energy obtained by using the first method are $8.923\times10^{29}$ erg and $2.059\times10^{30}$ erg, respectively, and the ones obtained by \EqAA~(\ref{eq:energy_pot_open}) is $8.603\times10^{29}$ erg and $2.090\times10^{30}$ erg, respectively, with relative errors of $3.586\%$ and $1.483\%$, respectively.

In the open field, all the magnetic field lines have one end rooted at the bottom surface and the other end extending up to infinity, so the field lines on the two side of the PIL run antiparallel, and in between a CS forms (see \FigAA~\ref{fig:open_slice}B and C). The current density outside CS is zero, and all free magnetic energy is stored through the CS (but \emph{not} in the CS). In the 3D prospective view of the open field as shown in \FigAA~\ref{fig:open_slice}, the structure of the CS is denoted by the red iso-surface of current density $J=0.34 \times 10^{-3}$ A m$^{-2}$. The central cross section of the open field and the profile of the current density along the central vertical line are also shown in \FigAA~\ref{fig:open_slice}. Since the magnetic field is discretized with a finite resolution, the current density in the CS is not infinite but rather changes with height; it first increases from nearly zero, reaches a peak value at a certain height, and then decrease towards zero again. Therefore, the maximum current density and its height can be obtained by calculating the current density in the CS. We consider that the CS of open field can be used as a proxy of the CS formed in the core field in our simulations, and particularly, the location of the maximum current density in the open field CS indicates the position where reconnection most likely starts to trigger an eruption, and the maximum current density itself should be related to the reconnection rate. Therefore, the two parameters; the peak value of current density of the open field CS and its height, are essential indicator of the strength of the CS, which is correlated with the strength of the eruption once being triggered. In order to calculate the current density of the open field CS, we first calculate the open field by the Green's function method and then use the second-order central difference to get the current density in the CS.

\begin{figure}[htbp]
	\centering
	\includegraphics[width=0.5\textwidth]{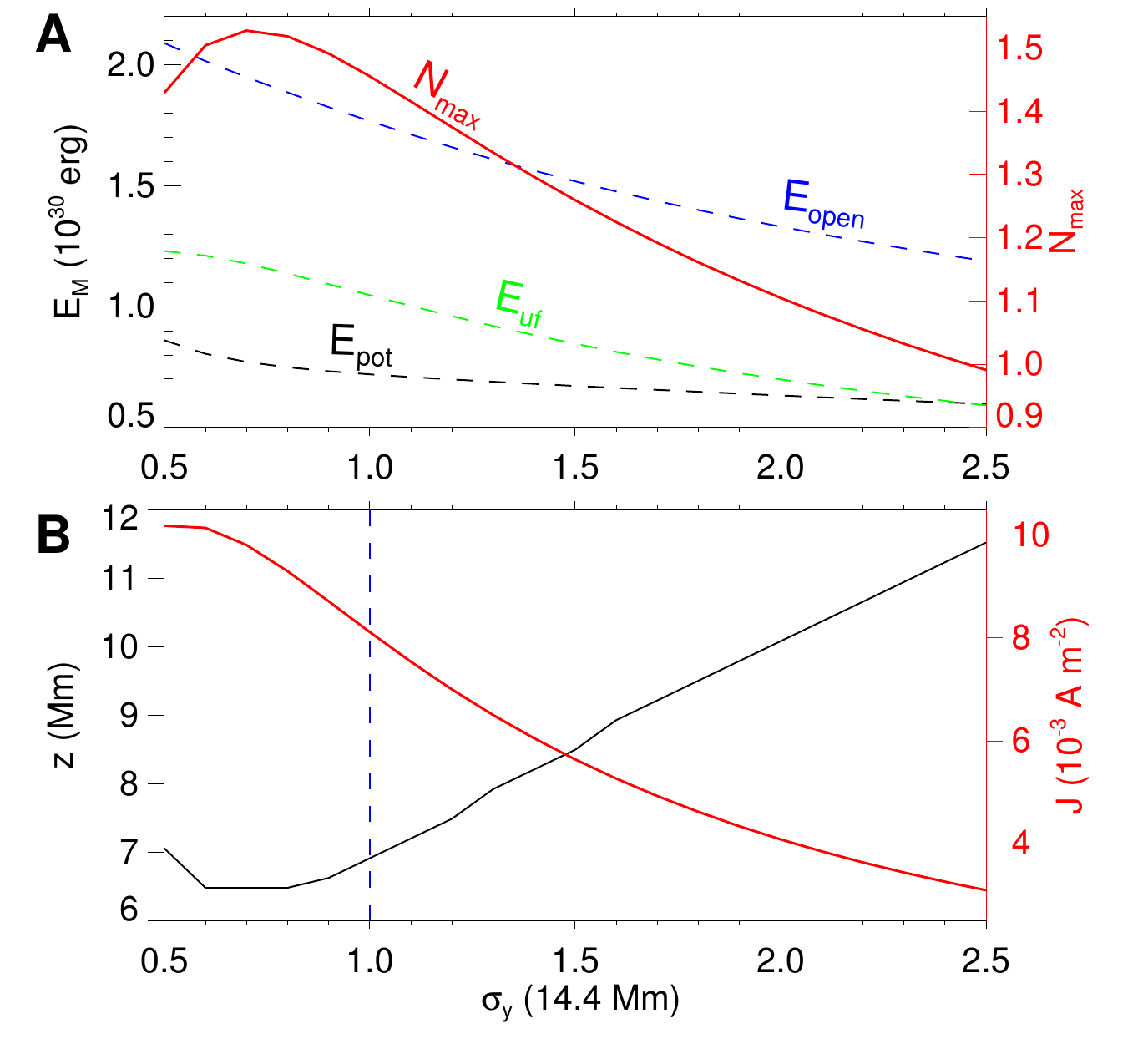}
	\caption{The magnetic energies, the non-potentiality $N_{\rm max}$ and distribution of CS varies with $\sigma_y$ (with $\sigma_x = 2.0$ and $y_c = 0.8$ fixed). \textbf{A} The black, blue and green curves represent potential energy, open field energy, and free magnetic energy, respectively. Red represents non-potentiality $N_{\rm max}$. \textbf{B} The black and red curves respectively represent the height and value of the maximum current density.}
	\label{fig:sigy_2001}
\end{figure}

\begin{figure}[htbp]
	\centering
	\includegraphics[width=0.5\textwidth]{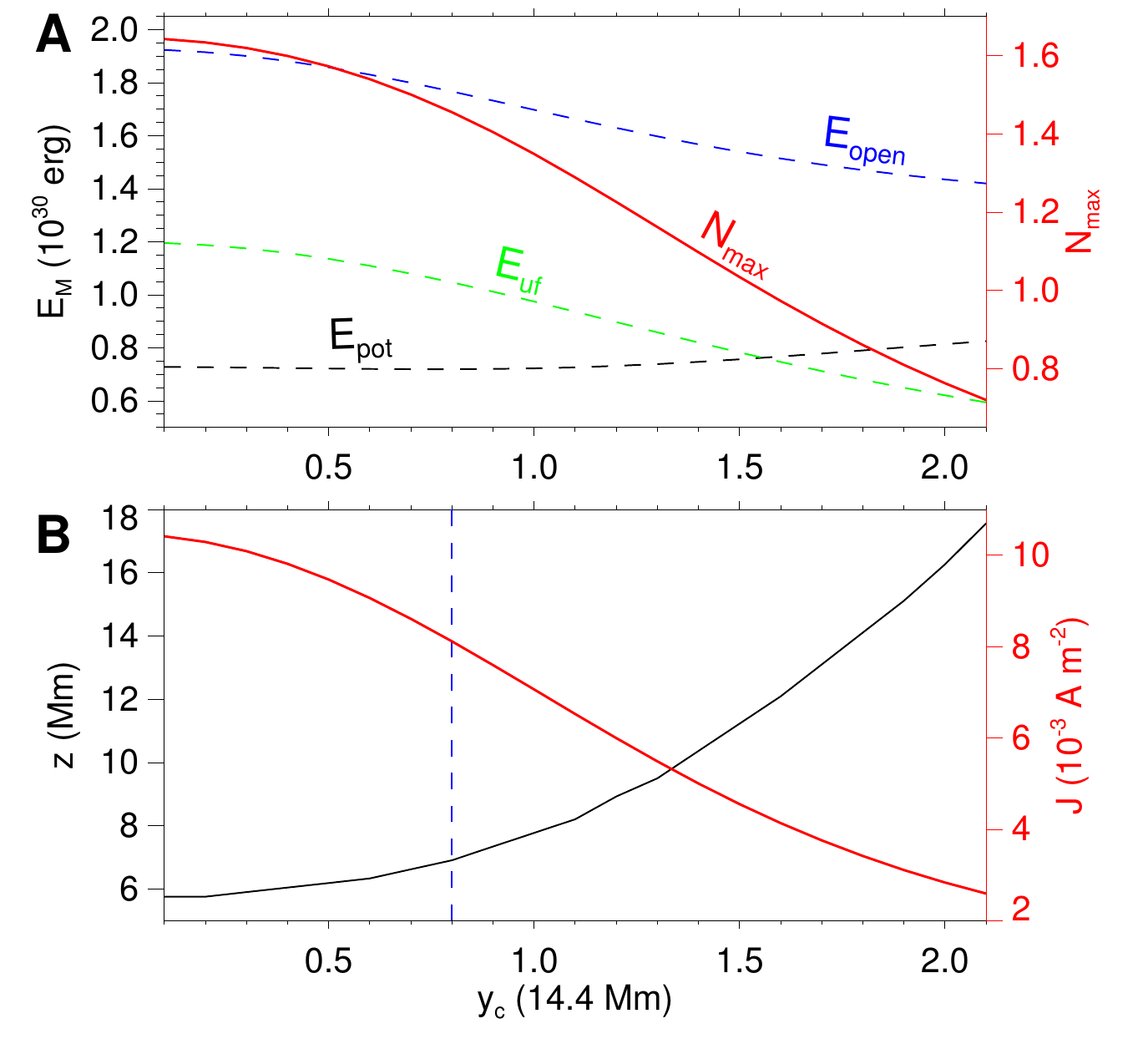}
	\caption{Same as \FigAA~\ref{fig:sigy_2001} but varies with $y_c$ (with $\sigma_x=2.0$ and $\sigma_y=1.0$ fixed).}
	\label{fig:yc_2001}
\end{figure}

\Fig~\ref{fig:sigy_2001} shows the result for different magnetograms specified by the parameter $\sigma_y$ changing from $0.5$ to $2.5$ with increment of $0.1$ (while $\sigma_x = 2.0$ and $y_c = 0.8$ are fixed), and \FigAA~\ref{fig:yc_2001} shows the results for magnetograms with parameter $y_c$ changing from $0.1$ to $2.1$ with increment of $0.1$ (with $\sigma_x = 2.0$ and $\sigma_y = 2.0$ fixed). Specifically, \FigAA~\ref{fig:sigy_2001}A shows the variation of different magnetic energies and the non-potentiality $N_{\rm max}$ with parameter $\sigma_y$. With the increase of $\sigma_y$, all energies decrease, and $N_{\rm max}$ also shows an overall decrease from about $1.5$ to below $1.0$, which clearly shows that, overall, the more elongated the magnetic polarity is, the more the upper limits of free energy and non-potentiality the magnetogram can reach (except that a too small $\sigma_y$ actually reduces $N_{\rm max}$, which is discussed below). \Fig~\ref{fig:yc_2001}A shows the variation of magnetic field energies and the non-potentiality $N_{\rm max}$ with parameter $y_c$. A similar decrease pattern is also seen (except that the potential energy increases mildly) with the increase of $y_c$, which shows that the closer the two magnetic poles are, the more free magnetic energy (and non-potentiality) the magnetogram can reach at most.

\begin{figure}[htbp]
	\centering
	\includegraphics[width=0.5\textwidth]{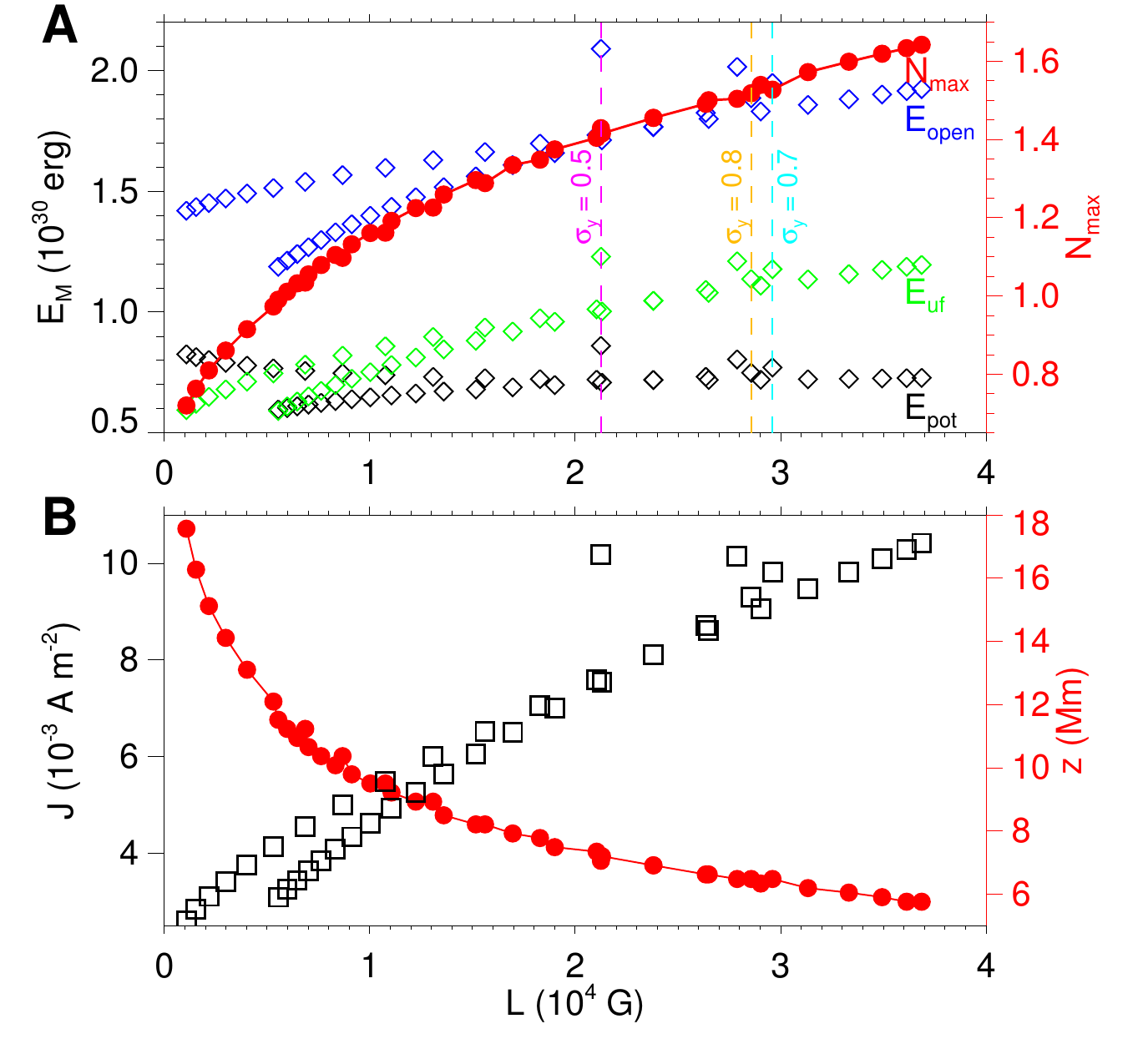}
	\caption{The relationship between the non-potentiality $N_{\rm max}$ and distribution of CS and the line integral of the $B_{z}$ gradient across the PIL on the magnetogram. \textbf{A} The diamond represents magnetic energies, and black, blue and green represent potential energy, open field energy, and free magnetic energy, respectively. The red solid circle represents non-potentiality $N_{\rm max}$. The three dashed vertical lines represent the values of three different parameters $\sigma_y$ (with $\sigma_x=2.0$ and $y_c=0.8$ fixed). Magenta, gold and clue represent $\sigma_y=0.5$, $0.7$ and $0.8$, respectively. \textbf{B} The black squares represents the maximum current density, and the red solid curves represents the height of the maximum current density.}
	\label{fig:L_versus_N}
\end{figure}

We note that, in \FigAA~\ref{fig:sigy_2001}A, $N_{\rm max}$ has a maximum at $\sigma_y \sim 0.8$ and a smaller $\sigma_y$ actually reduces $N_{\rm max}$. This is because as the polarity distance is fixed at $y_c = 0.8$, a too small value of $\sigma_y \le 0.8$, i.e., a too much concentration of the polarity in the $y$ direction, will actually make strong magnetic flux distributed farther away from the PIL, and thus the magnetic field gradient near the PIL is reduced. It hints that the gradient on the PIL is a more important indicator for $N_{\rm max}$. To confirm this, we calculate the line integral of gradient of the vertical field $B_z$ across the PIL for each magnetogram, named as $L$, which is defined by
\begin{equation}\label{eq:L}
	\begin{split}
		L=\int_{\rm PIL}\dfrac{\partial B_z}{\partial y}{\rm d}x.
	\end{split}
\end{equation}
\Fig~\ref{fig:L_versus_N}A shows the diagram of $L$ versus $N_{\rm max}$ for all the different sets of parameters shown in \FigsAA~\ref{fig:sigy_2001} and \ref{fig:yc_2001}. Strikingly, $N_{\rm max}$ increases monotonically with the increase of $L$, which conforms our argument. Hereafter we will refer to the parameter $L$ as simply the strength of the PIL. Note that $L$ for $\sigma_y = 0.5$ is less than that of $\sigma_y = 0.8$, and thus $N_{\rm max}$ of the former is less than that of the latter. We also note that with decreasing of $L$, $N_{\rm max}$ decreases more and more fast.

For the characteristic of the CS, \FigAA~\ref{fig:sigy_2001}B shows that with the increase of $\sigma_y$, the peak value of current density decreases and its height increases. This indicates that the more elongated the magnetic polarity is, the stronger the CS can form. In \FigAA~\ref{fig:yc_2001}B, with the increase of $y_c$, the maximum current value decreases and its height increases. This shows the closer the magnetic polarities are, the stronger the CS can form. Again, the PIL strength $L$ is crucial. As can be seen in \FigAA~\ref{fig:L_versus_N}, with the increase of $L$, the peak value of current density increases monotonically overall (except some points due the discretization errors), and the height decreases systematically, which means that the PIL strength is positively correlated with the strength of the CS.

\subsection{Settings of Numerical Model}
\label{sec:model}
For each magnetogram, we carried out an MHD simulation which begins with the potential field and is driven continually by a rotational flows at each magnetic polarity at the lower boundary, which creates magnetic shear along the PIL. The rotational flow is defined as
\begin{equation}\label{eq:dirven_speed}
	\begin{split}
		v_{x}=\dfrac{\partial \psi(B_{z})}{\partial y}; v_{y}=\dfrac{\partial \psi(B_{z})}{\partial x}
	\end{split}
\end{equation}
with $ \psi $ given by
\begin{equation}\label{eq:dirven_speed_psi}
	\begin{split}
		\psi = v_{0}B_{z}^{2}e^{-(B_{z}^{2}-B_{z, {\rm max}}^{2})/B_{z, {\rm max}}^{2}},
	\end{split}
\end{equation}
where $B_{z,{\rm max}}$ is the largest value of the photosphere $B_{z}$, and $v_{0}$ is a constant for scaling such that the maximum of the surface velocity is $4.4$~km~s$^{-1}$, close to the typical flow speed in the photosphere ($\sim$$1$~km~s$^{-1}$). The flow speed is smaller than the sound speed by two orders of magnitude and the local Alfv$\acute{\text{e}}$n speed by three orders, respectively, thus representing a quasi-static stress of the coronal magnetic field. The flow pattern has been shown in \FigsAA~\ref{fig:sigy_bz_four_figure} and \ref{fig:yc_bz_four_figure}. This is an incompressible, anti-clockwise rotational flow that does not change with time, and it will not modify the flux distribution at the bottom surface.

We numerically solve the full MHD equations with both coronal plasma pressure and solar gravity included, in a 3D Cartesian geometry by an advanced conservation element and solution element (CESE) method implemented on an adaptive mesh refinement (AMR) grid~\citep{Jiang2010,Feng2010,Jiang2016NC, jiang_fundamental_2021}. Since the controlling equations, the numerical code and essentially all the setting of initial and boundary conditions are the same as used in~\citet{jiang_fundamental_2021}, the readers are referred to that paper (in particular, the Method part) for more details. We note that no explicit resistivity is used in the magnetic induction equation, but magnetic reconnection can still occur due to numerical diffusion when a current layer is sufficiently narrow such that its width is close to the grid resolution.

The computational volume spans a Cartesian box of approximately $(-270,-270,0)$~Mm $\leqslant(x,y,z)\leqslant(270,270,540)$~Mm (where $z=0$ represents the solar surface). The volume is large enough such that the simulation runs can be stopped before the disturbance from the simulated eruption reaches any of these boundaries. The full volume is resolved by a block-structured grid with AMR in which the base resolution is $\Delta x=\Delta y=\Delta z=\Delta=2.88$~Mm, and the highest resolution of $\Delta =360$~km is used to capture the formation of CS and the subsequent reconnection.

We have run in total nine experiments for a selected set in the parameter space as listed in Table~\ref{tab:CASE_Expression}. Specifically, CASE \RNum{1} to CASE \RNum{4} represent $\sigma_y= 0.5$, $1.0$, $1.5$ and $2.0$, respectively (with fixed $\sigma_x= 2.0$ and $y_{c} = 0.8$). CASE \RNum{5} to CASE \RNum{8} represent $y_{c}= 0.1$, $0.6$, $1.1$ and $1.6$, respectively (with fixed $\sigma_x= 2.0$ and $\sigma_y= 1.0$). In order to verify the analysis of free magnetic energy maximum in \FigAA~\ref{fig:sigy_2001}, we set CASE \RNum{9}, in which $\sigma_x= 2.0$, $\sigma_y= 0.8$ and $y_{c}= 0.8$.

\begin{table}[htbp]
	\centering
	\begin{tabular}{lcccc}
		\hline\hline
		\multirow{2}{*}{Experiments}&\multicolumn{3}{c}{Expression}& \multirow{2}{*}{L ($10^4$~G)} \\ \cline{2-4}
		&$\sigma_x$ & $\sigma_y$ & $y_c$ \\ \cline{2-4}
		\hline
		CASE \RNum{1}  & 2.0 & 0.5 & 0.8&2.12  \\
		CASE \RNum{2}  & 2.0 & 1.0 & 0.8&2.38  \\
		CASE \RNum{3}  & 2.0 & 1.5 & 0.8&1.36  \\
		CASE \RNum{4}  & 2.0 & 2.0 & 0.8&0.83  \\
		CASE \RNum{5}  & 2.0 & 1.0 & 0.1&3.69  \\
		CASE \RNum{6}  & 2.0 & 1.0 & 0.6&2.90  \\
		CASE \RNum{7}  & 2.0 & 1.0 & 1.1&1.56  \\
		CASE \RNum{8}  & 2.0 & 1.0 & 1.6&0.53  \\
		CASE \RNum{9}  & 2.0 & 0.8 & 0.8&2.86  \\
		\hline
	\end{tabular}
	\caption{The parameters $\sigma_x$, $\sigma_y$ and $y_c$ that define the nine magnetograms for the simulated experiments. We also give the line integral (L) of the magnetic field gradient on the PIL for each magnetogram.}
	\label{tab:CASE_Expression}
\end{table}

\subsection{Simulation results}
\label{sec:results}
The BASIC scenario as demonstrated in \citet{jiang_fundamental_2021} is that, by continually shearing a bipolar coronal field, a CS forms slowly within the arcade and once reconnection sets in, the whole arcade explodes and forms a fast-ejecting magnetic flux rope (i.e., CME). Here we first show briefly an example of such process in CASE \RNum{1}, and then analyze the different runs based on the different magnetograms.

\begin{figure*}[htbp]
	\centering
	\includegraphics[width=0.85\textwidth]{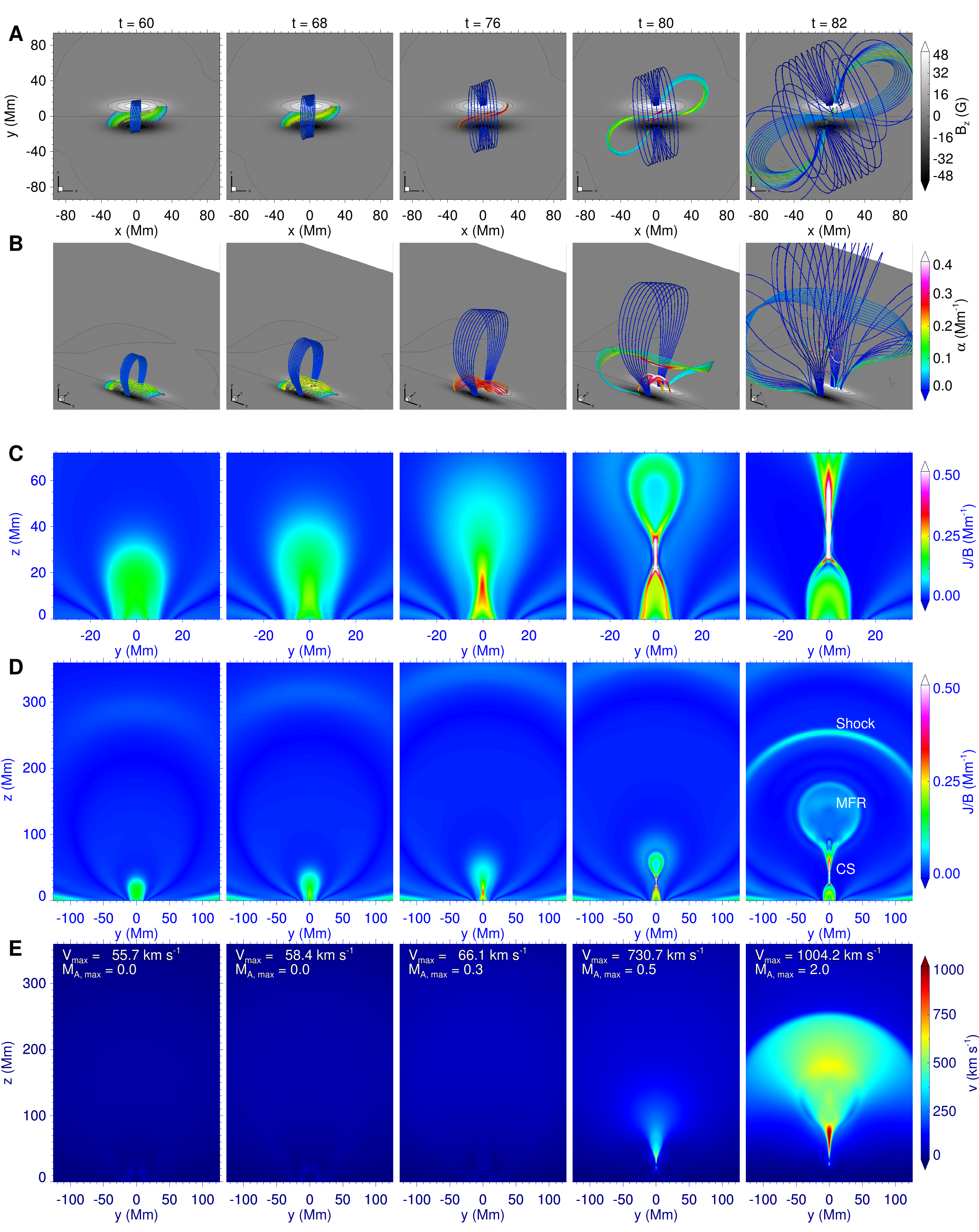}
	\caption{Evolution of magnetic field lines, electric currents, and velocity in the whole simulation process of CASE \RNum{1}. \textbf{A} Top view of magnetic field lines. The colored thick lines represent magnetic field lines and the colors denote the value of nonlinear force-free factor defined as $\alpha=\text{\textbf{J}} \cdot \text{\textbf{B}}/B^2$, which indicates how much the field lines are non-potential. The background shows the magnetic flux distribution on the bottom boundary (i.e., plane of $z = 0$), and contours of $B_z=(-48,-32,-16,0,16,32,48)$ G are shown. \textbf{B} 3D prospective view of the same field lines shown in panel \textbf{A}. \textbf{C} Current density $J$ normalized by magnetic field strength $B$ in vertical cross section (i.e., the $x = 0$ slice).	\textbf{D} Same as \textbf{C}, but with a large area. \textbf{E} Magnitudes of velocity. The largest velocity and Alfv$\acute{\text{e}}$nic Mach number are also denoted.}
	\label{fig:Evolution_CASE1}
\end{figure*}

\Fig~\ref{fig:Evolution_CASE1} and Supplementary Movie 1 show
evolution of magnetic field lines, current density, and velocity during the whole simulation process. The time unit is $\tau = 105$ s (all the times mentioned in this paper are expressed with the same time unit). After a period of surface flow driving, the magnetic field structure has evolved from the initial potential field to a configuration with a strong shear immediately above the PIL, where the current density is enhanced, while the envelope field is still current-free ($t \le 76$). It can be clearly seen that the entire magnetic field structure inflates during the energy injection stage, since the magnetic pressure of the core field increases gradually by the continuous shear. As a result, it stretches outward the envelope field, making the bipolar magnetic arcade tend to approach an open field configuration (\FigAA~\ref{fig:Evolution_CASE1}B). During this quasi-static evolution process, the current is squeezed from the volumetric distribution into a vertical, narrow layer extending above the PIL, forming a vertical CS (\FigAA~\ref{fig:Evolution_CASE1}C).

\begin{figure*}[htbp]
	\centering
	\includegraphics[width=0.8\textwidth]{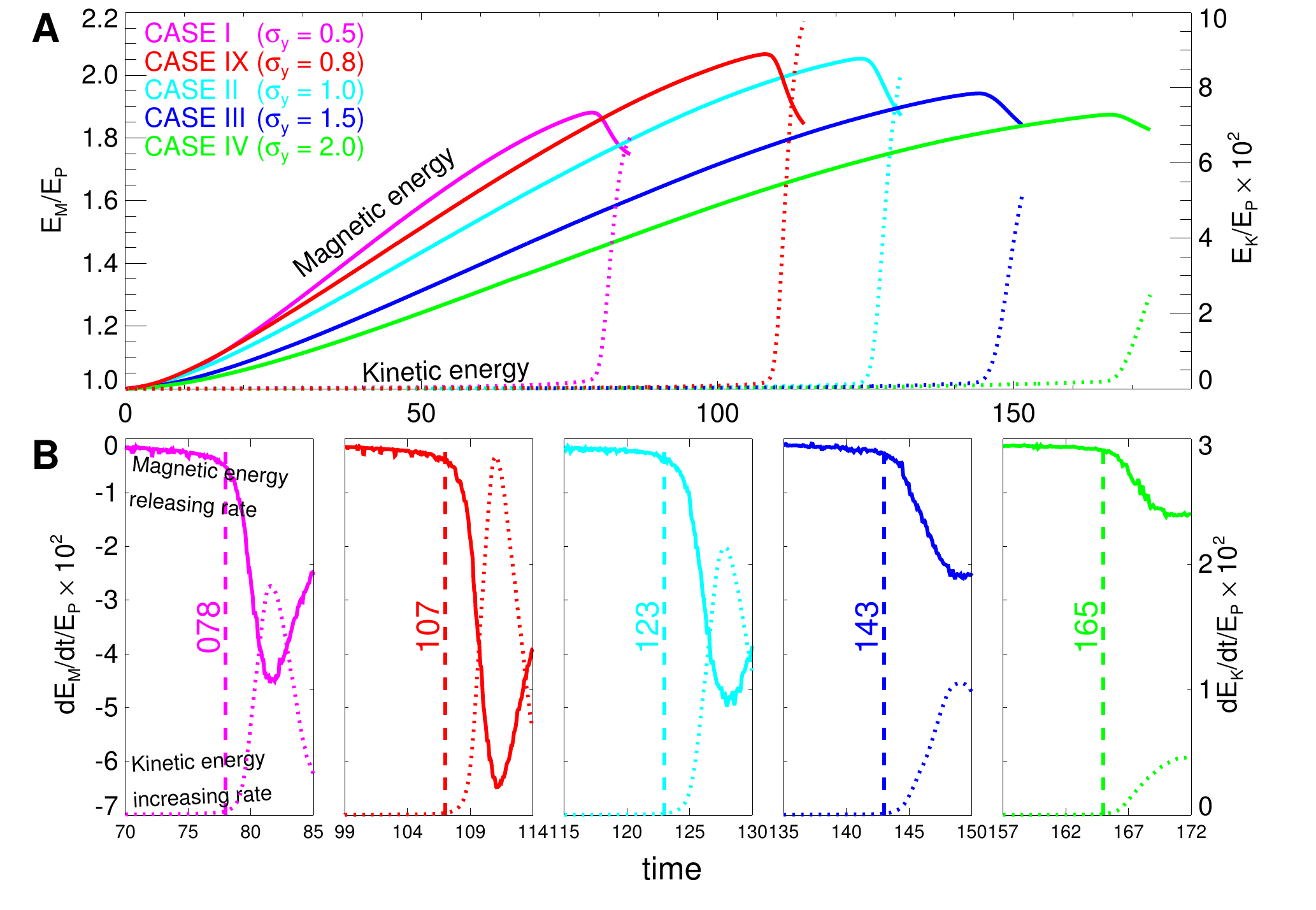}
	\caption{Temporal evolution of magnetic energy and kinetic energy in the simulations of different $\sigma_y$. The color of the lines represents five experiments, magenta, red, cyan, blue and green represent CASE \RNum{1}, \RNum{9}, \RNum{2}, \RNum{3}, \RNum{4}, respectively. \textbf{A} Evolution of magnetic energy $E_{\rm M}$ (solid lines) and kinetic energy $E_{\rm K}$ (dotted lines).  \textbf{B} Releasing rate of magnetic energy and increasing rate of kinetic energy of five experiments. The vertical dashed lines are shown for denoting the transition time of from pre-eruption to eruption. In the ordinates of all panels, the normalized unit is the potential field energy based on each magnetogram.}
	\label{fig:para_evol_SI_sigy}
\end{figure*}

As a critical point, when the thickness of the CS decreases down to the grid resolution, magnetic reconnection sets in and triggers an eruption. This transition from the pre-eruption to eruption onset is clearly manifested in the evolution of energies as shown in \FigAA~\ref{fig:para_evol_SI_sigy} (see the curves colored in magenta), which have a sharp transition at $t=78$. The kinetic energy increases impulsively to nearly $7\%$ original magnetic potential energy in a time duration of $\Delta t=5$. Meanwhile, the magnetic energy releases quickly during the eruption. The onset of the eruption can be more clearly shown by the time profiles of the magnetic energy release rate and the kinetic energy increase rate, and both of them have a sharp increase at the beginning of the eruption (\FigAA~\ref{fig:para_evol_SI_sigy}B).

\begin{figure*}[htbp]
	\centering
	\includegraphics[width=0.8\textwidth]{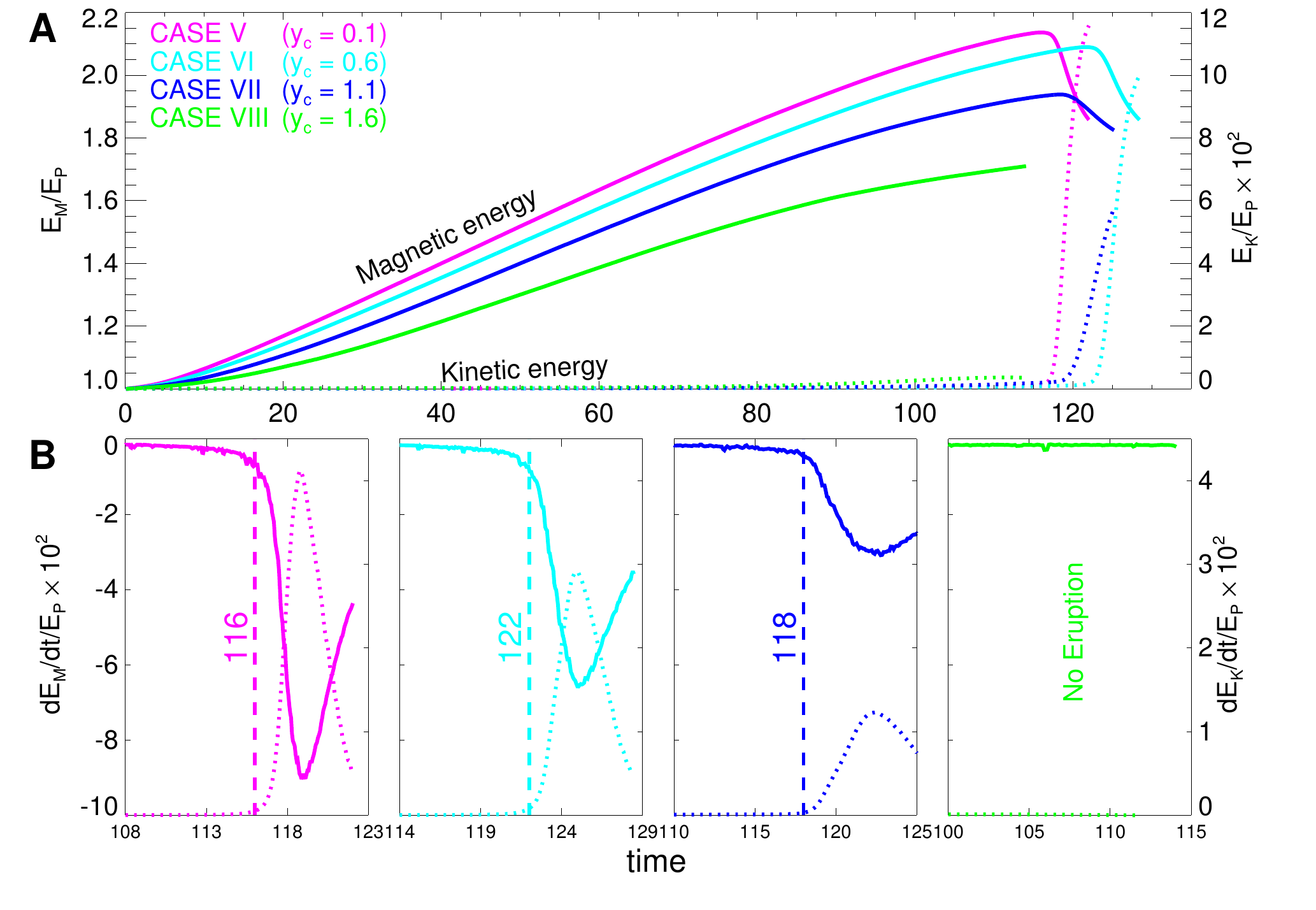}
	\caption{Same as \FigAA~\ref{fig:para_evol_SI_sigy} but with different $y_c$. The color of the lines represents four experiments, magenta, cyan, blue and green represent CASE \RNum{5}, \RNum{6}, \RNum{7}, \RNum{8}, respectively. Note that CASE \RNum{8} did not produce an eruption.}
	\label{fig:para_evol_SI_yc}
\end{figure*}

\begin{figure*}[htbp]
	\centering
	\includegraphics[width=0.8\textwidth]{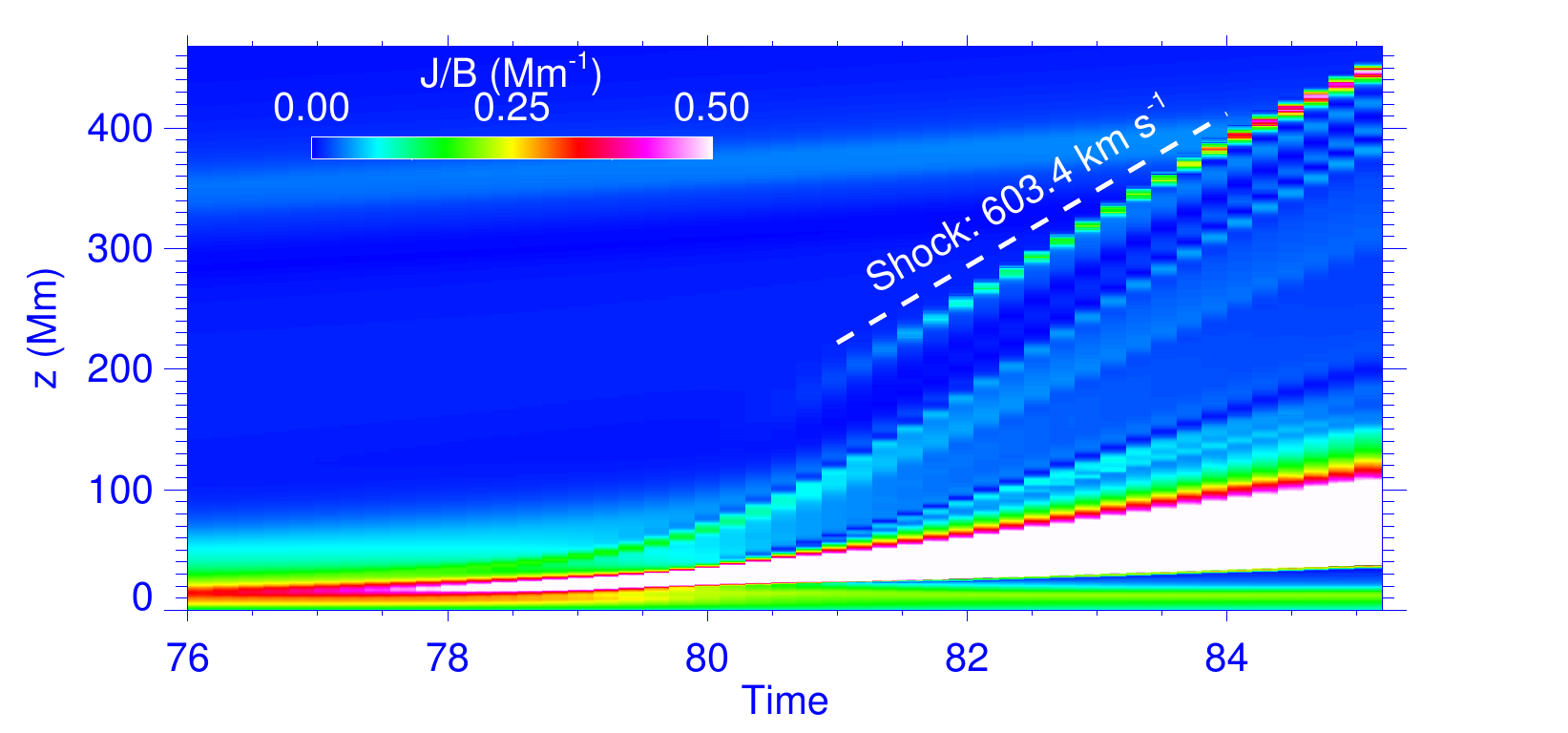}
	\caption{A time stack map of the current distribution around $x,y = 0$ in CASE \RNum{1}, which can reveal the evolution speed of the CME.}
	\label{fig:velocity}
\end{figure*}

With the onset of reconnection, a plasmoid (i.e., MFR in 3D) originates from the tip of the CS and rises quickly, leaving behind a cusp structure separating the reconnected, post-flare loops from the un-reconnected field (\FigAA~\ref{fig:Evolution_CASE1}C and D). The plasmoid expands quickly and meanwhile, an arc-shaped fast magnetosonic shock is formed in front of the plasmoid. All of these evolving structures are proof of the typical coronal magnetic eruption leading to CME. The shock marks the front edge of the CME, and its average speed is about $603.4$ km s$^{-1}$ (\FigsAA~\ref{fig:Evolution_CASE1}D and \ref{fig:velocity}).

\Figs~\ref{fig:para_evol_SI_sigy} and \ref{fig:para_evol_SI_yc} show the temporal evolution of magnetic and kinetic energies (and their changing rate) of all the cases from CASE \RNum{1} to \RNum{9}. Note that in each CASE, the energies are normalized by the corresponding potential field energy, i.e., the magnetic energy at $t=0$. Overall, all the different runs (except CASE \RNum{8}) show a similar evolution pattern: magnetic energy first increases monotonically for a long time, approaching the open field energy, while the kinetic energy remains to be a very low level; at a critical point, the magnetic energy begins to decrease rapidly along with an impulsive rise of the kinetic energy and the evolutions of the two energies are closely correlated in time, which indicates that the free magnetic energy is released to accelerate the plasma. In Supplementary Movies 2 and 3, we show the evolution of current density on the central cross section for all the cases. As can be seen, they all follow the same BASIC scenario; that is, first a CS forms during the magnetic energy increasing phase and then reconnection sets in and triggers eruption. Therefore, these different runs demonstrate the robustness of the BASIC mechanism.

\begin{figure}[htbp]
	\centering
	\includegraphics[width=0.42\textwidth]{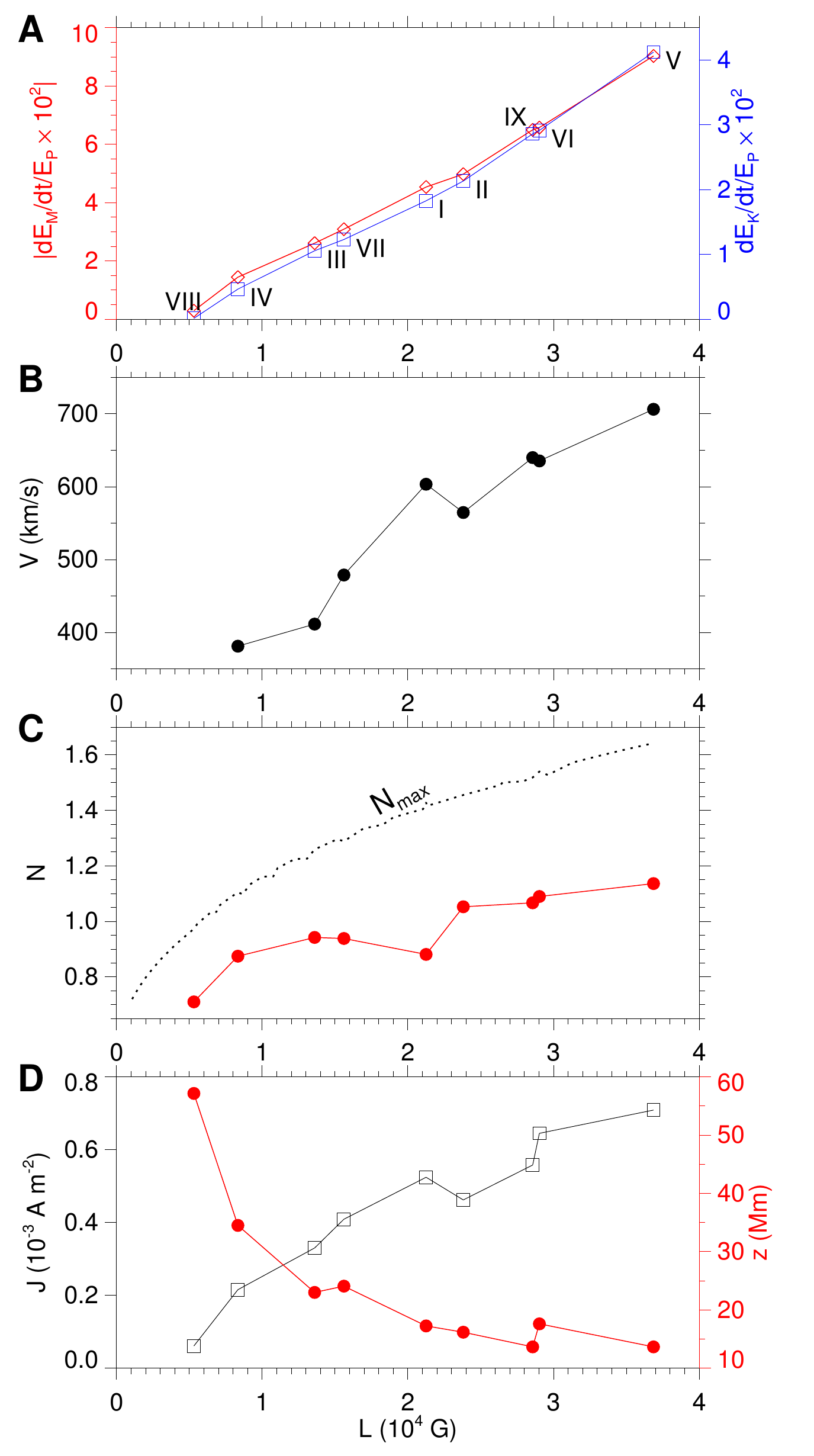}
	\caption{The intensity of the eruption, as quantified by (\textbf{A}) the peak value of the kinetic energy increasing rate and the magnetic energy releasing rate as well as (\textbf{B}) the largest speed of the leading edge of the CME, depend on $L$. \textbf{C} non-potentiality $N$ at the eruption onset time, the dotted line represents the theoretical calculation value, same as \FigAA~\ref{fig:L_versus_N}. \textbf{D} The location of CS and the maximum current density in CS at the eruption onset time.}
	\label{fig:explosive_intensity}
\end{figure}

Nevertheless, the magnitudes, or intensities, of the eruptions in the different cases are different, as can be seen by comparing the energy conversion rates during the impulsive phase. For example, in the five experiments with increasing $\sigma_y$ as shown in \FigAA~\ref{fig:para_evol_SI_sigy}, the maximum release rate of magnetic energy increases first, reaching the largest at $\sigma_y=0.8$ and then decreases with higher $\sigma_y$, which is exactly consistent with the dependence of $N_{\rm max}$ on $\sigma_y$ as shown in \FigAA~\ref{fig:sigy_2001}. Again, \FigAA~\ref{fig:para_evol_SI_yc} shows that with the increase of $y_c$, the eruption intensity decreases, consistent with the dependence of $N_{\rm max}$ on $y_c$ as shown in \FigAA~\ref{fig:yc_2001}. In \FigAA~\ref{fig:explosive_intensity}A and B, we further show how the strength of the PIL, i.e., $L$, is related with the intensity of the eruption, as quantified by the peak values of the kinetic energy increasing rate and the magnetic energy releasing rate as well as the speed of the leading edge of the CME (i.e., the shock). It clearly shows that the eruption intensity is correlated positively with the PIL strength. And in \FigAA~\ref{fig:explosive_intensity}C, we show the non-potentiality $N$ at the onset time of the eruption, as compared with the corresponding $N_{\rm max}$. The non-potentiality increases overall with increase of the PIL strength, consistent with (but a bit slower than) that of the $N_{\rm max}$. Interestingly, we note that the non-potentiality at the eruption onset is mostly close to one, which strikingly agrees with the statistical study of observed active regions \citep{moore_limit_2012}, and hints that the BASIC mechanism is responsible for those eruptions.

\begin{figure*}[htbp]
	\centering
	\includegraphics[width=0.8\textwidth]{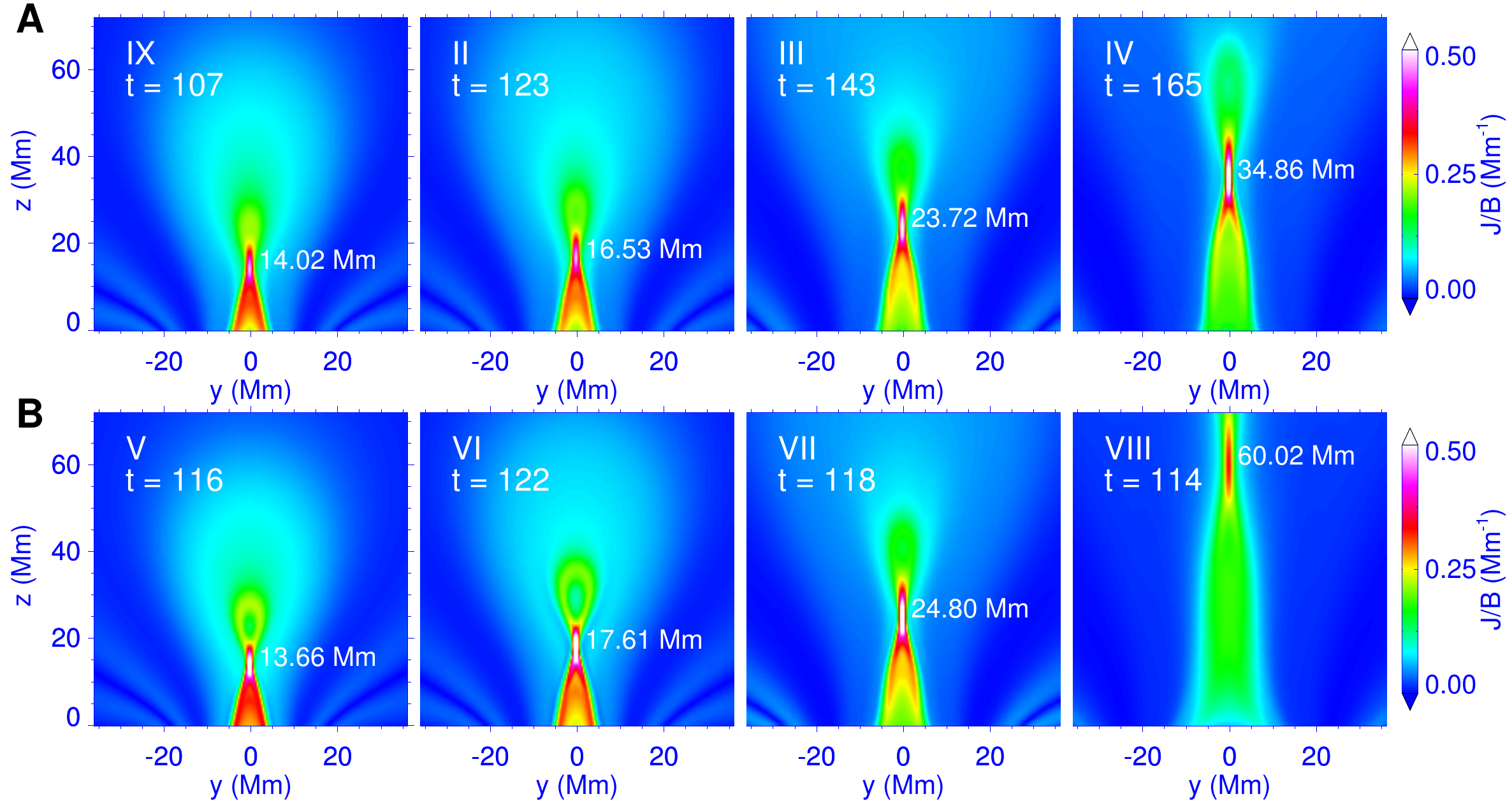}
	\caption{The vertical cross section of the current density $J$ (normalized by magnetic field strength $B$) at the eruption onset time for the eight experiments. From left to right and top to bottom are CASE \RNum{9}, \RNum{2}, \RNum{3}, \RNum{4}, \RNum{5}, \RNum{6}, \RNum{7}, and \RNum{8}. The location of the CS and the maximum current density in CS are denoted on each panel. Noticed that there is no eruption for CASE \RNum{8}.}
	\label{fig:cs_form_sigy_yc}
\end{figure*}

\Fig~\ref{fig:cs_form_sigy_yc} shows the central vertical cross section of current density $J$ (normalized by magnetic field strength $B$) at time immediately close to the eruption onset time for the different experiments. The location of the CS and the maximum current density in the CS are also shown in \FigAA~\ref{fig:explosive_intensity}D. These results are consistent with the calculation of the open field in Section~\ref{sec:magnetic field analyze}. Furthermore, we can find that the lower the position of the CS and the higher the current density are, the greater the explosive intensity is (\FigAA~\ref{fig:explosive_intensity}).

The only case in our experiments that did not reach an eruption is CASE \RNum{8}. This is because during the quasi-static shearing process, the field expands fast in the later phase (e.g., $t>100$) and strongly presses the numerical boundaries before a CS is formed (or before a sufficient amount of free magnetic energy is accumulated to approach a open field), therefore we have stopped the simulation run to avoid a too much influence from the numerical boundaries on the results. Ideally, with a larger computational box, a free expansion of the field driven by the surface shearing motion will create a CS, but its height is too large (and the current density is too low) to trigger an efficient eruption.

It is interesting that the ratio $E_{\rm open}/E_{\rm pot}$ of around $1.7$ is apparently a threshold to start an eruption. Such a similar ratio has been found in Amari's simulations on flux rope formation and instability~\citep{amari_twisted_2000,Amari2003A,Amari2003B}. Actually, for all the different distributions of magnetic flux as we have considered in this paper, the lowest ratio of $E_{\rm open}/E_{\rm pot}$ is $1.7$ (see \FigAA~\ref{fig:L_versus_N}A, where $N_{\rm max}=E_{\rm open}/E_{\rm pot}-1$). The reason why $E_{\rm open}/E_{\rm pot}$ of $1.7$ appears to be a threshold to start an eruption is that in our scenario the CS can only form (and thus to trigger an eruption) when the magnetic field is sufficiently sheared such that its energy is close to the open field energy. Furthermore, as we have analyzed, the $E_{\rm open}/E_{\rm pot}$ (or $N_{\rm max}$) should be large enough to let the CS to form at a low height and with a large current density, such that the reconnection can be efficient to produce an eruption. Otherwise, if the $E_{\rm open}/E_{\rm pot}$ is too small, the free energy that can be attained is small, and the CS will form at a too large height (and with too low current density) to trigger an efficient eruption, as our experiment CASE VIII shows. The smaller the ratio $E_{\rm open}/E_{\rm pot}$ is, the higher the CS will form, and the smaller the free energy can be reached, and thus the less efficiently the eruption can be produced. For example, in the extreme case when $E_{\rm open}/E_{\rm pot}$ reaches it a lower limit of $E_{\rm open}/E_{\rm pot}=1$, which corresponds to the flux distribution with two opposite polarities being infinitely far away from each other, it cannot produce any eruption no matter how large the polarities are rotated.  

Although using a different scenario (i.e., first a flux rope is created by surface converging and/or cancellation and then the flux rope runs into instability to initiate an eruption), Amari's simulations also show the same behavior that their eruption occurs only when the magnetic energy is close to the open field energy. Thus, both of our and Amari's simulations using a single bipolar flux distribution suggest that the ratio $E_{\rm open}/E_{\rm pot}$ should be larger than $1.7$ to start an eruption. Amari's simulations consist of two important phases of energizing. Their first phase is the same as ours, i.e., by rotating the polarities to inject free energy into the field. The key difference is that in \citet{amari_twisted_2000,Amari2003A,Amari2003B} simulations the rotation is stopped before a CS is formed; then, in the second phase, they modified the flux content by opposite flux emerging~\citep{amari_twisted_2000} or surface diffusion~\citep{Amari2003A}, and/or modified the flux distribution by surface converging flow~\citep{Amari2003B} at the bottom boundary. In this phase, the magnetic topology will be changed from a sheared arcade to a flux rope, through the slow reconnection near the bottom boundary. More importantly, the corresponding open field of the evolving magnetic flux distribution will change, being faster than that of the total magnetic energy, and can eventually lead to an eruption when the total magnetic energy is close to the open field energy.

Finally, we note that the eruption onset times of the different experiments are different. This is related to the magnetic energy injection rate, which depends on the surface flow distribution. As shown in \FigAA~\ref{fig:para_evol_SI_sigy}, the magnetic energy injection rate decreases when $\sigma_y$ increases, and thus the eruption onset time is systematically postponed, because longer time is needed for free magnetic energy accumulation.

\section{Conclusions}
\label{sec:conclusions}

It has long been known that major solar eruptions mostly occur in active regions with strongly-sheared and strong-gradient PIL. There is no doubt that a strong magnetic shear is critical for producing eruption, since it is directly related to the degree of non-potentiality of the field. However, it lacks an explanation why the flux distribution with a high-gradient PIL is favorable for eruption. In this paper we provide such a physics explanation, for the first time, basing on the BASIC mechanism with different photospheric magnetic flux distributions, i.e., magnetograms, by combining theoretical analysis and numerical simulation. The BASIC mechanism refers to a simple and efficient scenario in which a internal CS can form slowly in a gradually sheared bipolar field and reconnection of the CS triggers and drives the eruption~\citep{jiang_fundamental_2021}.

In principle, two requirements are essential to initiate a major eruption by the BASIC mechanism, and they are closely related to each other. One is to accumulate a sufficient amount of free magnetic energy to power a major eruption, namely, the field should be sufficiently non-potential. The other is to build up a strong CS in the bipolar core field, that is, a CS with a high current density and formed at a low height, such that reconnection can release magnetic energy efficiently. Focusing on these two key elements, we set up a series of magnetograms with equaling unsigned flux but different flux distributions and first analyzed the open fields corresponding to these magnetograms. This is because the open field sets an upper limit for the energy that a sheared bipolar field can store as well as the intensity of CS that the field can form. By calculating the largest non-potentiality $N_{\rm max}$ and the peak current density based on the open field, we find that magnetogram having a stronger PIL can contain more non-potentiality $N_{\rm max}$ and can form stronger CS, which indicates the capability to initiate a larger eruption by the BASIC mechanism. Furthermore, we find that the strength of the PIL, named as $L$, can be quantified well by a line integral of gradient of the vertical field $B_z$ across the PIL, which should be valuable in future studies for flare forecast based on magnetograms.

Then we selected nine representative magnetograms to conduct MHD simulations. All of the numerical experiments exhibit the same evolution pattern; magnetic energy first increases monotonically for a long time as driven by the boundary rotational flow, and in the duration the kinetic energy remains within a very low level, indicating that the system is a quasi-static evolution process. Then at a critical point when the thickness of the CS decreases down to the grid resolution, reconnection sets in and triggers an eruption, during which the magnetic energy decrease rapidly along with an fast rise of the kinetic energy. A overall comparison of the eruptions in the different cases shows a strong correlation of the eruption intensity with the strength of the PIL. Specifically, with the increase of the PIL strength, both the non-potentiality of the field and the strength of the CS at the eruption onset increase, and consequently, the eruption intensity increases, which confirms the two key conditions in the BASIC mechanism.

In summary, through the combined study of theoretical analysis and numerical simulations, we demonstrated that the bipolar field with magnetogram of strong PIL can hold more non-potentiality and can form stronger internal CS, which are key to the initiation of strong eruption. This is the physical reason why magnetic field with a strong PIL is capable of producing major eruption. Our study demonstrates the robustness of the BASIC mechanism on the one hand, and discloses the physics reason why a long and strong-gradient PIL is favorable for major eruption on the other hand.

\begin{acknowledgements}
This work is jointly supported by National Natural Science Foundation of China (NSFC 41822404, 41731067, 41574170, and 41531073), and Shenzhen Technology Project JCYJ20190806142609035. 
The computational work was carried out on TianHe-1(A), National Supercomputer Center in Tianjin, China. 
\end{acknowledgements}

\bibliographystyle{aa}
\bibliography{all}

\begin{thebibliography}{}
\expandafter\ifx\csname natexlab\endcsname\relax\def\natexlab#1{#1}\fi
\providecommand{\url}[1]{\href{#1}{#1}}
\providecommand{\dodoi}[1]{doi:~\href{http://doi.org/#1}{\nolinkurl{#1}}}
\providecommand{\doeprint}[1]{\href{http://ascl.net/#1}{\nolinkurl{http://ascl.net/#1}}}
\providecommand{\doarXiv}[1]{\href{https://arxiv.org/abs/#1}{\nolinkurl{https://arxiv.org/abs/#1}}}

\bibitem[{Aly(1991)}]{aly_how_1991}
Aly, J.~J. 1991, \apj, 375, L61

\bibitem[{{Amari} {et~al.}(2018){Amari}, {Canou}, {Aly}, {Delyon}, \&
  {Alauzet}}]{Amari2018}
{Amari}, T., {Canou}, A., {Aly}, J.-J., {Delyon}, F., \& {Alauzet}, F. 2018,
  \nat, 554, 211

\bibitem[{{Amari} {et~al.}(2003{\natexlab{a}}){Amari}, {Luciani}, {Aly},
  {Mikic}, \& {Linker}}]{Amari2003A}
{Amari}, T., {Luciani}, J.~F., {Aly}, J.~J., {Mikic}, Z., \& {Linker}, J.
  2003{\natexlab{a}}, \apj, 585, 1073

\bibitem[{{Amari} {et~al.}(2003{\natexlab{b}}){Amari}, {Luciani}, {Aly},
  {Mikic}, \& {Linker}}]{Amari2003B}
{Amari}, T., {Luciani}, J.~F., {Aly}, J.~J., {Mikic}, Z., \& {Linker}, J.
  2003{\natexlab{b}}, \apj, 595, 1231

\bibitem[{{Amari} {et~al.}(1996){Amari}, {Luciani}, {Aly}, \&
  {Tagger}}]{Amari1996}
{Amari}, T., {Luciani}, J.~F., {Aly}, J.~J., \& {Tagger}, M. 1996, \aap, 306,
  913

\bibitem[{Amari {et~al.}(2000)Amari, Luciani, Mikic, \&
  Linker}]{amari_twisted_2000}
Amari, T., Luciani, J.~F., Mikic, Z., \& Linker, J. 2000, ApJ, 529, L49

\bibitem[{{Antiochos} {et~al.}(1999){Antiochos}, {DeVore}, \&
  {Klimchuk}}]{Antiochos1999}
{Antiochos}, S.~K., {DeVore}, C.~R., \& {Klimchuk}, J.~A. 1999, \apj, 510, 485

\bibitem[{{Aulanier}(2014)}]{Aulanier2014}
{Aulanier}, G. 2014, in IAU Symposium, Vol. 300, IAU Symposium, ed.
  B.~{Schmieder}, J.-M. {Malherbe}, \& S.~T. {Wu}, 184--196

\bibitem[{{Aulanier} {et~al.}(2000){Aulanier}, {DeLuca}, {Antiochos},
  {McMullen}, \& {Golub}}]{Aulanier2000}
{Aulanier}, G., {DeLuca}, E.~E., {Antiochos}, S.~K., {McMullen}, R.~A., \&
  {Golub}, L. 2000, \apj, 540, 1126

\bibitem[{{Aulanier} {et~al.}(2010){Aulanier}, {T{\"o}r{\"o}k}, {D{\'e}moulin},
  \& {DeLuca}}]{Aulanier2010}
{Aulanier}, G., {T{\"o}r{\"o}k}, T., {D{\'e}moulin}, P., \& {DeLuca}, E.~E.
  2010, \apj, 708, 314

\bibitem[{{Chen}(2011)}]{ChenP2011}
{Chen}, P.~F. 2011, Living Rev. Solar Phys., 8, 1

\bibitem[{{Choe} \& {Lee}(1996)}]{Choe1996}
{Choe}, G.~S. \& {Lee}, L.~C. 1996, \apj, 472, 360

\bibitem[{Falconer(2003)}]{falconer_measure_2003}
Falconer, D.~A. 2003, J. Geophys. Res., 108, 1380

\bibitem[{{Falconer} {et~al.}(2002){Falconer}, {Moore}, \&
  {Gary}}]{Falconer2002}
{Falconer}, D.~A., {Moore}, R.~L., \& {Gary}, G.~A. 2002, \apj, 569, 1016

\bibitem[{{Fan} \& {Gibson}(2007)}]{Fan2007}
{Fan}, Y. \& {Gibson}, S.~E. 2007, \apj, 668, 1232

\bibitem[{{Feng} {et~al.}(2010){Feng}, {Yang}, {Xiang}, {Wu}, {Zhou}, \&
  {Zhong}}]{Feng2010}
{Feng}, X.~S., {Yang}, L.~P., {Xiang}, C.~Q., {et~al.} 2010, \apj, 723, 300

\bibitem[{{Forbes} {et~al.}(2006){Forbes}, {Linker}, {Chen}, {Cid}, {K{\'o}ta},
  {Lee}, {Mann}, {Miki{\'c}}, {Potgieter}, {Schmidt}, {Siscoe}, {Vainio},
  {Antiochos}, \& {Riley}}]{Forbes2006}
{Forbes}, T.~G., {Linker}, J.~A., {Chen}, J., {et~al.} 2006, \ssr, 123, 251

\bibitem[{{Janvier} {et~al.}(2015){Janvier}, {Aulanier}, \&
  {D{\'e}moulin}}]{Janvier2015}
{Janvier}, M., {Aulanier}, G., \& {D{\'e}moulin}, P. 2015, \solphys, 290, 3425

\bibitem[{Jiang {et~al.}(2021)Jiang, Feng, Liu, Yan, Hu, Moore, Duan, Cui, Zuo,
  Wang, \& Wei}]{jiang_fundamental_2021}
Jiang, C., Feng, X., Liu, R., {et~al.} 2021, Nat Astron, 10.1038/s41550-021-01414-z

\bibitem[{Jiang {et~al.}(2010)Jiang, Feng, Zhang, \& Zhong}]{Jiang2010}
Jiang, C.~W., Feng, X.~S., Zhang, J., \& Zhong, D.~K. 2010, \solphys, 267, 463

\bibitem[{{Jiang} {et~al.}(2016){Jiang}, {Wu}, {Feng}, \& {Hu}}]{Jiang2016NC}
{Jiang}, C.~W., {Wu}, S.~T., {Feng}, X.~S., \& {Hu}, Q. 2016, Nature Comm., 7,
  11522

\bibitem[{{Kliem} \& {T{\"o}r{\"o}k}(2006)}]{Kliem2006}
{Kliem}, B. \& {T{\"o}r{\"o}k}, T. 2006, Physical Review Letters, 96, 255002

\bibitem[{Künzel(1959)}]{kunzel_flare-haufigkeit_1959}
Künzel, H. 1959, Astr. Nachr.; AN, 285, 271

\bibitem[{Lynch {et~al.}(2008)Lynch, Antiochos, DeVore, Luhmann, \&
  Zurbuchen}]{Lynch2008}
Lynch, B.~J., Antiochos, S.~K., DeVore, C.~R., Luhmann, J.~G., \& Zurbuchen,
  T.~H. 2008, \apj, 683, 1192

\bibitem[{Mikic \& Linker(1994)}]{mikic_disruption_1994}
Mikic, Z. \& Linker, J.~A. 1994, \apj, 430, 898

\bibitem[{Moore {et~al.}(2012)Moore, Falconer, \& Sterling}]{moore_limit_2012}
Moore, R.~L., Falconer, D.~A., \& Sterling, A.~C. 2012, \apj, 750, 24

\bibitem[{{Moore} \& {Labonte}(1980)}]{Moore1980}
{Moore}, R.~L. \& {Labonte}, B.~J. 1980, in IAU Symposium, Vol.~91, Solar and
  Interplanetary Dynamics, ed. M.~{Dryer} \& E.~{Tandberg-Hanssen}, 207--210

\bibitem[{{Moore} \& {Roumeliotis}(1992)}]{Moore1992}
{Moore}, R.~L. \& {Roumeliotis}, G. 1992, in Lecture Notes in Physics, Berlin
  Springer Verlag, Vol. 399, IAU Colloq. 133: Eruptive Solar Flares, ed.
  Z.~{Svestka}, B.~V. {Jackson}, \& M.~E. {Machado}, 69

\bibitem[{{Moore} {et~al.}(2001){Moore}, {Sterling}, {Hudson}, \&
  {Lemen}}]{Moore2001}
{Moore}, R.~L., {Sterling}, A.~C., {Hudson}, H.~S., \& {Lemen}, J.~R. 2001,
  \apj, 552, 833

\bibitem[{Schmieder {et~al.}(2013)Schmieder, Démoulin, \&
  Aulanier}]{Schmieder2013}
Schmieder, B., Démoulin, P., \& Aulanier, G. 2013, Advances in Space Research,
  51, 1967

\bibitem[{Schrijver(2007)}]{schrijver_characteristic_2007}
Schrijver, C.~J. 2007, \apj, 655, L117

\bibitem[{{Shibata} \& {Magara}(2011)}]{Shibata2011}
{Shibata}, K. \& {Magara}, T. 2011, Living Rev. Solar Phys., 8, 6

\bibitem[{Sturrock(1991)}]{sturrock_maximum_1991}
Sturrock, P.~A. 1991, \apj, 380, 5

\bibitem[{Sun {et~al.}(2015)Sun, Bobra, Hoeksema, Liu, Li, Shen, Couvidat,
  Norton, \& Fisher}]{SunX2015}
Sun, X., Bobra, M.~G., Hoeksema, J.~T., {et~al.} 2015, \apjl, 804, L28

\bibitem[{Toriumi \& Wang(2019)}]{Toriumi2019}
Toriumi, S. \& Wang, H. 2019, Living Rev. Solar Phys., 16, 3

\bibitem[{{T{\"o}r{\"o}k} \& {Kliem}(2005)}]{Torok2005}
{T{\"o}r{\"o}k}, T. \& {Kliem}, B. 2005, \apjl, 630, L97

\bibitem[{{Wyper} {et~al.}(2017){Wyper}, {Antiochos}, \& {DeVore}}]{Wyper2017}
{Wyper}, P.~F., {Antiochos}, S.~K., \& {DeVore}, C.~R. 2017, \nat, 544, 452

\end{thebibliography}

%
%

\end{document}